\documentclass[pra,aps,amsmath,amssymb,onecolumn]{revtex4}
\usepackage{epsfig}
\usepackage{bm}
\usepackage{color}
\begin{document}

\preprint{}

%%% to see comments, use first version, to hide them, use second one
\newcommand{\stef}[2]{$\blacktriangleright${\sc round #1:}{\em #2}$\blacktriangleleft$}

\title{Numerical analysis of relaxation times of multiple quantum coherences 
in the system with a large number of spins}

\author{ S.I.Doronin}
\email{s.i.doronin@gmail.com}
\author{E.B.Fel'dman}
\email{efeldman@icp.ac.ru}
\author{A.I.Zenchuk}
\email{zenchuk@itp.ac.ru}

\affiliation{Institute of Problems of  Chemical Physics of the Russian Academy of Sciences, Chernogolovka, Moscow Region, 142432,Russia}
%{\color{red} ,,,, }
\begin{abstract}
 We study the decay of multiple quantum (MQ) NMR coherences 
in systems with the large number of equivalent spins. As being created on the preparation period of MQ NMR experiment, they decay due to the dipole-dipole interactions (DDI) on the evolution period of this experiment.  It is shown that the relaxation time decreases with the increase in  MQ coherence order { (according to the known results)} and in the number of spins. We also consider  the modified preparation period of  MQ NMR experiment (G.A.Alvarez, D.Suter, PRL {\bf 104}, 230403 (2010)) concatenating the short evolution periods under the secular DDI Hamiltonian (the perturbation) with the evolution period under the non-secular averaged two-spin/two-quantum Hamiltonian. The influence of the perturbation on the decoherence rate is investigated for the systems consisting of 200-600 equivalent spins.
    
\end{abstract}


\maketitle

\section{Introduction}

Multiple quantum (MQ) coherences are quite suitable for investigations of  the dependence of 
the relaxation time on the size of the quantum system \cite{KS1,KS2,AS,CCCR}. This problem is closely connected to the estimations of the decoherence time  as an important parameter for the quantum information systems. A simplest model  of the quantum register formed by the highly correlated spins can be created in MQ NMR experiments \cite{BMGP}. { Some models of quantum registers} consisting of up to 4900 qubits were studied experimentally \cite{KS2}.
{ The theoretical methods (for example, ref.\cite{FF}) describing the experiments \cite{KS1} are phenomenological ones and the development of theoretical and numerical approaches from "the first principles" are fully justified. 
 At the same time, numerical methods of MQ NMR dynamics allow us, generally speaking, to investigate systems consisting of not more than twenty spins \cite{DFGM}.  Some progress in the 
study of the larger systems (up to 40 spins) 
is achieved due to the special techniques based on the  Chebyshev polynomial expansion \cite{DRKH,ZCAPCDRV} and on the phenomenon of  quantum parallelism \cite{ADLP}. The new perspectives are opened by MQ NMR in systems of equivalent spins where the special method has been worked out \cite{DFFZ1,DFFZ2} allowing one to investigate MQ NMR dynamics of hundreds of spins and even more.} Such systems
of equivalent spins can be created in a nanopore compound placed in a strong external magnetic field if the nanopores are filled with a gas of spin-carring molecules (atoms) \cite{BKHWW,FR}. Since the characteristic time of the molecular diffusion is much less than the spin flip-flop time \cite{BKHWW,FR},  the dipole-dipole interactions (DDI)  of spins are averaged ({ but not to zero}) and the residual DDI can be described by the { single} coupling constant. As a result, all spins can be considered as equivalent ones, which significantly simplifies the numerical simulations.

The above method  can be applied to the investigation of the  decay of  MQ NMR coherence intensities of different orders caused by the secular DDI in systems containing hundreds of spins. In the simplest case this decay occurs  on the evolution period of  the MQ NMR experiment \cite{KS1}. 
However, the MQ NMR experiment can be modified, for instance, using a different set of pulses on the preparation period, as in Ref.\cite{AS}. In the later case, the decay takes place on the preparation period.

In this paper we study the decay  of  MQ NMR  coherence intensities created  in systems with a large number of equivalent spins. The paper is organized as follows. The general description of different  MQ NMR experiments is given in Sec.\ref{Section:gen}.  The theory and the numerical simulation of the decay of  MQ NMR coherence intensities in different MQ NMR experiments  is developed in Sec.\ref{Section:theory}. 
The conservation law associated with considered models is derived in Sec.\ref{Section:discussion}. We briefly summarize our results in  concluding Sec.\ref{Section:conclusion}.

%%%%%%%%%%%
\section{The MQ NMR experiments in a system of equivalent spins}
\label{Section:gen}
\begin{figure}
   \epsfig{file=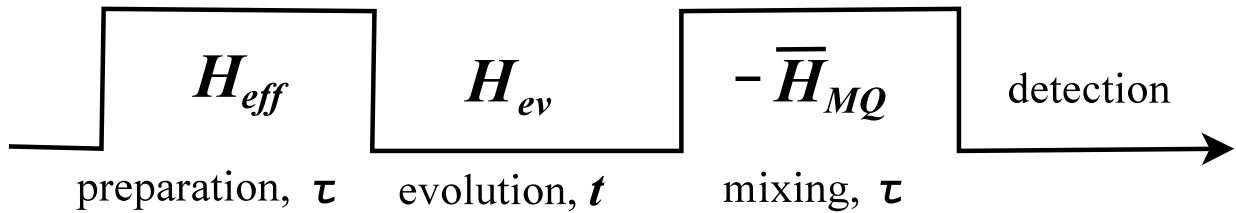
%figexp1
   ,scale=0.6,angle=0}
  \caption{The basic scheme of the  MQ NMR experiment. The Hamiltonian $H_{eff}$ (Eq.(\ref{H_eff})), $H_{ev}$ (Eq.(\ref{H_ev}))and $\bar H_{MQ}$ (Eq.(\ref{H_MQ})) govern the spin dynamics on the appropriate period of the MQ NMR experiment}
  \label{Fig:1}
\end{figure}

The  MQ NMR experiment consists of four distinct periods of time (Fig.\ref{Fig:1}): preparation ($\tau$), evolution ($t$), mixing ($\tau$) and  detection.
% ($t_2$).
%
%%
\paragraph{Preparation period.}
 The spin system is irradiated by the proper multipulse sequence 
%(eight-pulse cycles in the standard experiments, \cite{BMGP}) 
on the preparation period. As a result, the anisotropic DDI of nuclear spins in the external magnetic field, $\vec B$, (directed along the axis $z$)  oscillates rapidly. In the rotating reference frame \cite{G}, the dynamics of spin system is described by the effective Hamiltonian $H_{eff}$. 
{ We consider two types of pulse sequences on the preparation period. 
The first one is the standard pulse sequence resulting in 
the averaged non-secular two-spin/two-quantum Hamiltonian, describing  the MQ NMR dynamics on the preparation period of the standard MQ NMR experiment
\cite{WSWP,BMGP}, i.e. $H_{eff}\equiv H_{MQ}$:
\begin{eqnarray}\label{H}
&&
H_{MQ}=H^{(+2)}+H^{(-2)},
\\
\label{Hpm}
&&
H^{(\pm 2)} =-\frac{1}{2} \sum_{j<k} D_{jk} I^{\pm}_jI^{\pm}_k,
\end{eqnarray}
%of Eq.(\ref{HMQ}) 
Here $D_{jk} =\gamma^2 \hbar(1-3\cos^2\theta_{jk})/(2r^3_{jk})$ is the coupling constant between spins $j$ and $k$, $\gamma$ is the gyromagnetic ratio, $r_{jk}$ is the distance between spins $j$ and $k$,  $\theta_{jk}$ is the angle between the vectors $\vec r_{jk}$ and $\vec B$  and $I^\pm_j=I_{jx}\pm i I_{jy}$ are the raising and lowering operators of  spin $j$.
  The second type of pulse sequences is introduced in Ref.\cite{AS}, where 
a modification of the preparation period of the  MQ NMR experiment was suggested. In this case,
the  preparation period consists of the cycles of the duration $\tau_c$ and { each} cycle concatenates the short evolution period $\tau_{dz}$ under the perturbation Hamiltonian $ H_{dz}$ (which is responsible for the secular DDI \cite{G}),
% of Eq.(\ref{Hdz}) 
\begin{eqnarray}\label{H_dz0}
H_{dz} = \sum_{j<k} D_{jk} (2 I_{jz} I_{kz} -I_{jx}I_{kx}-I_{jy}I_{ky})
\end{eqnarray}
with the evolution period $\tau_{MQ}$ under the ideal MQ Hamiltonian $H_{MQ}$  (\ref{H}).
Thus, $\tau_c=\tau_{dz}+\tau_{MQ}$. }
Introducing the relative strength $p=\tau_{dz}/\tau_{c}$ ($0\le p \le 1$) of the perturbation one can find that the resulting evolution can be described  by the effective Hamiltonian $H_{eff}$ given by the following equation \cite{AS}:
\begin{eqnarray}\label{H_eff0}
H_{eff}(p)=(1-p) H_{MQ} + p H_{dz}.
\end{eqnarray}
%\begin{eqnarray}\label{H_eff}
%H_{eff}(p)=(1-p)\bar H_{MQ} + p \bar H_{dz}.
%\end{eqnarray}
Let  the preparation period consist of $K$ (a big number)  cycles of the duration $\tau_c$, so that one can  introduce the   parameter $\tau=K\tau_c$.  Note, that $H_{eff}(0)\equiv H_{MQ}$, which means that the standard preparation period, used, for instance, in ref. \cite{BMGP}, is a particular case of the described modification.

Hereafter we study  the MQ NMR dynamics of equivalent spins. Such a case can be realized, for instance,  by the dipolar coupling spins in a nanopore where the Hamiltonian (\ref{H}) is averaged ({ but not to zero}) by the fast molecular diffusion \cite{BKHWW,FR}. The Hamiltonians  $H_{MQ}$ and $H_{dz}$ with the averaged coupling constant $D$  ($D\tau_c\ll 1$) can be rewritten as follows \cite{DFFZ1,DFFZ2}:
\begin{eqnarray}\label{HMQ}\label{H_MQ}
&&\bar H_{MQ} = -\frac{D}{4}\{(I^{+})^2 +(I^{-})^2\},\\
\label{Hdz}\label{H_dz}
&&\bar H_{dz}=\frac{D}{2} \{3 I^2_z-I^2\},
\end{eqnarray}
where $I^\pm=\sum\limits_{j=1}^N I^\pm_j$ ($N$ is the number of spins), 
%The Hamiltonian $H_{dz}$ in Eq.(\ref{H_eff}) is responsible for the secular DDI \cite{FR}. In the system  of the %equivalent spins it reads:
%\begin{eqnarray}\label{Hdz}\label{H_dz}
%\bar H_{dz}=\frac{D}{2} \{3 I^2_z-I^2\},
%\end{eqnarray}
 $I_z=\sum\limits_{j=1}^N I_{zj}$ and the operator $I^2$ is the square of the total spin angular momentum. 
Thus Eq.(\ref{H_eff0}) must be replaced with the following one
\begin{eqnarray}\label{H_eff}
H_{eff}(p)=(1-p) \bar H_{MQ} + p \bar H_{dz},
\end{eqnarray}
which is valid 
for the system of equivalent spins.

%%%%%%%%%%%%%%%%
\paragraph{Evolution period.}

%The decay of MQ NMR coherences is caused by the Hamiltonian $\bar H_{dz}$ appearing in
Let the spin system  be governed by the following general Hamiltonian
 $H_{ev}$ on the evolution period:
\begin{eqnarray}\label{H_ev}
H_{ev}=(1-\theta(p)) \bar H_{dz} + \Delta I_z, \;\;\theta=\left\{\begin{array}{ll}
0,& p=0\cr
1,&p>0
\end{array}\right..
\end{eqnarray}
%The case $p=0$ in Sec.\ref{Section:first} or $\alpha=0$ in Sec.\ref{Section:second}.
 The offset $\Delta$ encodes MQ NMR coherences of different orders, { see  below, Eq. (\ref{Iz}) and ref.\cite{BMGP}}.

%%%%%%%%%%%%%%%
\paragraph{Mixing period.}

The spin system on the mixing period is governed by the Hamiltonian $-\bar H_{MQ}$ in all experiments considered in this paper. 

We emphasize that the decay of MQ NMR coherences is caused by the Hamiltonian $\bar H_{dz}$ appearing either  on the evolution period (if $p=0$ in Eqs.(\ref{H_eff},\ref{H_ev})) or on the preparation period 
(if $p\neq 0$ in eqs.(\ref{H_eff},\ref{H_ev})).
These two cases are considered separately in   Secs.\ref{Section:first} and \ref{Section:second} respectively.

%%%%%%%%%%%%
\section{The decay of MQ NMR coherence intensities  caused by the secular DDI 
%on the evolution period of MQ NMR experiments
}

\label{Section:theory}
\subsection{The decay of MQ NMR coherences in MQ NMR experiments of Ref. \cite{BMGP}}
\label{Section:first}
We consider the time evolution of the coherences in  MQ NMR experiments with the standard preparation period \cite{BMGP}. For this purpose we take $p=0$ in Eqs.(\ref{H_eff}) and (\ref{H_ev}), which read:
\begin{eqnarray}
H_{eff}=\bar H_{MQ},\;\;H_{ev} = \bar H_{dz} + \Delta I_z,
\end{eqnarray} 
 so that the coherence decay occurs on the evolution period. 
In order to investigate the MQ NMR dynamics of the system one should find the density matrix $\rho(\tau)$  on the preparation period solving the Liouville evolution equation \cite{G}
\begin{eqnarray}\label{L}
i \frac{d\rho(\tau)}{d\tau}=[\bar H_{MQ},\rho(\tau)]
\end{eqnarray}
with the initial thermodynamic equilibrium state $\rho(0)=I_z$ in the high temperature approximation \cite{G}.
Taking into account the pointed information about the Hamiltonians on the different periods of  MQ NMR experiment  one can write the expression for the longitudinal polarization $\langle I_z\rangle(\tau,t)$ after the mixing period of  MQ NMR experiment  (Fig.\ref{Fig:1})
as follows:
\begin{eqnarray}\label{I_z}
&&
\langle I_z\rangle(\tau,t)={\mbox{Tr}}\{U^+(\tau) e^{-i\Delta t I_z} e^{-i \bar H_{dz} t} U(\tau) 
\times\\\nonumber
&&
I_z U^+(\tau) 
e^{i\Delta t I_z} e^{i \bar H_{dz} t}U(\tau) I_z\}=\\\nonumber
&&
{\mbox{Tr}}\{ e^{-i\Delta t I_z} e^{-i \bar H_{dz} t} \rho(\tau)  e^{i \bar H_{dz} t} 
e^{i\Delta t I_z} \rho(\tau)\},
\end{eqnarray}
where $\rho(\tau)=U(\tau) I_z U^+(\tau)$ is the solution to Eq.(\ref{L}) and $U(\tau)=\exp(-i\bar H_{MQ} \tau)$. It is convenient to expand the spin density matrix $\rho(\tau)$ in the series as follows
\begin{eqnarray}
\rho(\tau) =\sum_k\rho_k(\tau),
\end{eqnarray}
{ where $\rho_k(\tau)$ is the contribution to $\rho(\tau)$ from MQ coherence of the $k$th order and satisfies the following commutation relation \cite{FL}:}
\begin{eqnarray}\label{phi_com}
e^{-i  \Delta tI_z} \rho_k e^{i \Delta t I_z} =e^{-i k \Delta t } \rho_k 
.
\end{eqnarray}
%{  $(\rho_k)_{ij}=\rho_{i+k,j}\delta_{ij}$, so that $[I_z,\rho_k]=k\rho_k$. 
%Here $\delta_{ij}$ is the Kronecker simbol.}
Then Eq.(\ref{I_z}) reads
\begin{eqnarray}\label{Iz}
\langle I_z\rangle(\tau,t)=\sum_k e^{-ik\Delta t}{\mbox{Tr}}\{
e^{-i\bar H_{dz} t} \rho_k(\tau)e^{i\bar H_{dz} t}\rho_{-k}(\tau)
\}.
\end{eqnarray}
Eq.(\ref{Iz}) defines the intensity $J_k(\tau,t)$ of  the MQ NMR coherence of order $k$ as follows:
\begin{eqnarray}\label{J_n}
J_k(\tau,t)={\mbox{Tr}}\{e^{-i\bar H_{dz} t} \rho_k(\tau)e^{i\bar H_{dz} t}\rho_{-k}(\tau)
\}.
\end{eqnarray}
{ In analogy to the autocorrelation function for the decay of the transverse magnetization \cite{G},
Eq.(\ref{J_n}) reveals  the decay of MQ NMR coherences due to the secular DDI on the evolution period. }
Since we consider a system of equivalent spins, numerical simulation of Eq.(\ref{J_n}) may be   simplified allowing one  to perform calculations in the systems with the large number of spins. This happens due to the  commutation relation  
\begin{eqnarray}
[\bar H_{MQ},I^2]=0,
\end{eqnarray} 
which suggests us to use  the basis of common eigenvectors of $I^2$ and $I_z$ \cite{DFFZ1}. It was shown  \cite{DFFZ1,DFFZ2} that  the Hamiltonians $\bar H_{MQ}$, $\bar H_{dz}$ and the density matrix $\rho(\tau)$ 
{ for the system  of equivalent spins} have a block structure. { For instance, $\bar H_{MQ}={\mbox{diag}}\{ \bar H^{\frac{N}{2}}_{MQ} ,\bar H^{\frac{N}{2}-1}_{MQ},
\dots ,\bar H^{\frac{N}{2}-\left[\frac{N}{2}\right]}_{MQ}\}$, ($[a]$ is an integer part of $a$) }. These blocks correspond to different total spin numbers $S=N/2,N/2-1,\dots, N/2-[N/2]$  \cite{LL}. All blocks are degenerated and their degeneracy $n_N(S)$ is determined as follows \cite{LL}:
\begin{eqnarray}
n_N(S)=\frac{N!(2 S+1)}{(\frac{N}{2} + S +1)!(\frac{N}{2} -S)!},\;\;0\le S\le \frac{N}{2}.
\end{eqnarray}
Thus, the problem is reduced to the set of analogous problems of lower dimensions. The intensities of MQ NMR  coherences $J_{k,S}(\tau,t)$ can be calculated for all blocks. Then the observable intensities $J_k(\tau,t)$ ($-N\le k\le N$) are following \cite{DFFZ1}:
\begin{eqnarray}
J_k(\tau,t) =\sum_Sn_N(S) J_{k,S}(\tau,t).
\end{eqnarray}
The results of numerical simulations of these intensities are { represented} below.
%in Sec.\ref{Section:numerics}.

\subsubsection{The numerical simulations}
\label{Section:num_first}

{ 
We study  the dynamics of MQ NMR coherence intensities in the nanopore filled with the spin-carring particles  numerically. Let us emphasize one more time that we are dealing with the highly symmetrical model where any two spins interact with the same constant of DDI
because the diffusion characteristic time in the nanopore  is much shorter then the spin flip-flop time \cite{BKHWW,FR}. This fact   simplifies the numerical calculations significantly since all particles  are "nearest neighbors" in this model and we  consider interactions among all of them.
}

\begin{figure}
   \epsfig{file=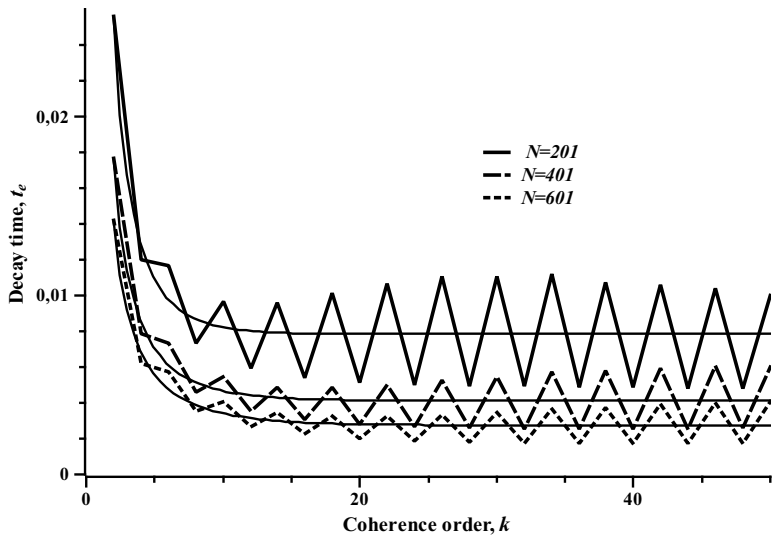
%figexp12
   ,scale=0.6,angle=270}
  \caption{The decay time as a function of the coherence order for the spin systems with $N=201$, 401 and 601. }
  \label{Fig:exp12}
\end{figure}

{ Our calculations showed \cite{DFFZ1} that MQ  NMR coherence intensities are  quickly oscillating functions. }
For this reason we follow the strategy of Ref.  \cite{DFFZ1} and consider the averaged intensities
\begin{eqnarray}\label{bJ_n}
&&
\bar J_k(\bar t) =\frac{1}{2 T}\int_{\tau_0}^{\tau_0+2 T} J_k(\bar \tau,\bar t)d\bar \tau,\\\nonumber
&&
\tau_0=31, \;\;T=2 \pi/\lambda_{min} =  2\pi/\sqrt{3}.
\end{eqnarray}
where $\bar \tau=D\tau$  and $\bar t=D t$ are the dimensionless times associated with the preparation and evolution periods respectively, { $\lambda_{min}$ is the minimal eigenvalue of the $\bar H_{MQ}$ Hamiltonian. This value belongs to the block $\bar H_{MQ}^{\frac{3}{2}}$} of the Hamiltonian \cite{DFFZ1}. 
{ The choice of $\tau_0$ is motivated by the requirement that  the coherences of all possible
orders have appeared and one can think that the quasi-stationary distribution of the intensities is realized, which has been  verified in ref.\cite{DFFZ1}.
The averaging is performed over two maximal periods $T$ of oscilations, which is taken
from the requirement that the increase of the averaging interval does not
change $\bar J_k$
 \cite{DFFZ1,DFFZ2}. }
The averaged intensities decay with the time  $\bar t$ of the evolution period. The time moments $t_e$ 
(such that $\bar J_k(0)/\bar J_k(\bar t)|_{\bar t=t_e}=e$ for an arbitrary $k$) versus MQ coherence order  in  systems  with 201, 401 and 601 spins are shown in Fig.\ref{Fig:exp12}.
%Times of decreasing the amplitudes of MQ NMR coherences versus MQ coherences orders in systems with 201, 401 and 601 spins %are presented in Fig.\ref{Fig:exp12}. 
We can see from this figure that 
\begin{enumerate}
\item
 MQ NMR coherence  decay times decrease with the increase   in the number of spins;
\item
MQ NMR coherence  decay times  decrease with the increase in their order.
\end{enumerate} 
The times $t_e(k)$ of the  decay of  MQ NMR coherences of order $k>0$ can be approximated by the hyperbolic cotangent, as it is shown in  Fig.\ref{Fig:exp12}:
\begin{eqnarray}\label{coth}
t_e(k) = a_1\coth(a_2 k + a_3),
\end{eqnarray}
where parameters $a_1,a_2,a_3$ may be found by the least square method:
\begin{eqnarray}\nonumber
t_e(k)&=&0.0078 \coth(0.1966 k-0.0758 ),\;\;N=201,\\\nonumber
t_e(k)&=&
0.0041 \coth(0.1441 k-0.0495 ),\;\;N=401,\\\nonumber
t_e(k)&=&0.0027 \coth(0.1144 k-0.0324 ),\;\;N=601.
\end{eqnarray}
\begin{figure}
  \epsfig{file=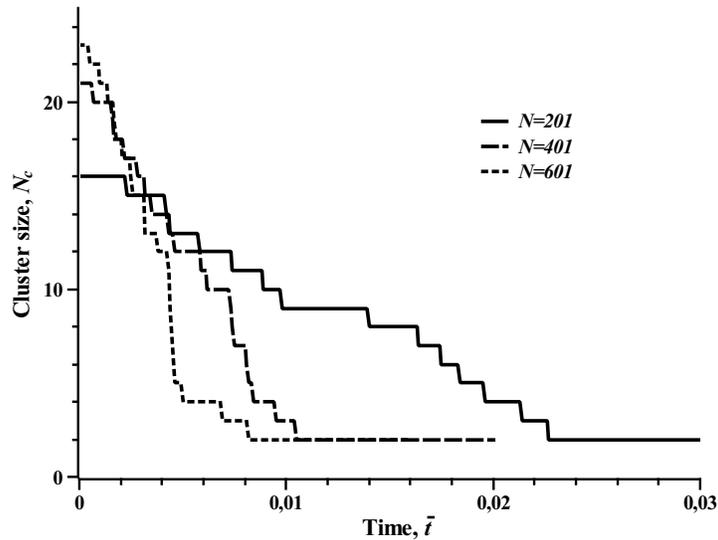
%figexpev13
   ,scale=0.6,angle=270}
  \caption{The evolution of the "cluster size" $N_c$; $\bar t$ is the dimensionless evolution time }
  \label{Fig:exp13}
\end{figure}
{ The approximation given by Eq.(\ref{coth}) shows that the  decay rate of the high order coherence intensities is almost independent on their order. This happens because MQ coherence phases (acquired during the evolution period) are approximately proportional to their order, see Eq. (\ref{phi_com}). As a result, the rates of  MQ coherence decays increase with their order and the decay time $t_e(k)$ is   $\sim 1/k$. Thus, for the high order coherences, we have $t_e(k)/t_e(k+1)\to 1$, i.e. the decay times of the $k$th and $(k+1)$th coherences are almost the same, which is reflected in Eq.(\ref{coth}). Regarding the zero-order coherence, its intensity $\bar J_0$ does not decay owing to the commutation relation $[\bar H_{dz},\rho_0]=0$, which follows from the fact that both $\bar H_{dz}$ (\ref{H_dz}) and $\rho_0$ are diagonal in the chosen basis.}

{ 
It is worthwile  to note that the dynamics of the multi-spin
cluster growth during the  evolution of the solid spin system considered, for instance, in \cite{BMGP,KS3} is essentially different in comparison with that  in the system of equivalent spins. The matter is that  only the strongly interacting spins are joined in the clusters initially, usually the nearest neighbors in the crystal lattice \cite{BMGP,KS3}. After that, the next neighbors become involved in the cluster and so on. Thus more and more remote spins become embedded in the cluster with time.  As a result, it becomes possible to observe the growth  of the multi-spin clusters in  MQ NMR experiments \cite{BMGP,KS3}. However, the dynamics of the spin clusters is quite different in the high symmetrical spin system such as the system of equivalent spins. All spins are "nearest neighbors" in this case, so that the spin cluster consisting of all $N$ spins is formed much more quickly during the time interval $\sim 1/D$, where $D$ is the constant of DDI, which is the same for any two spins. It becomes hard to follow the process of the cluster growth in the high symmetrical system of equivalent spins, unlike the solids \cite{BMGP,KS3}. Nevertheless, there is some reorganization of the spin cluster diring the evolution,  when the system is irradiated by the multipulse sequence \cite{WSWP,BMGP}, resulting to the high order MQ coherences.

Now let us turn to the decay of MQ NMR coherences. We 
 consider the "cluster" of MQ NMR coherences as a family of such coherences whose intensities exceed some fixed value $J_{min}$, say, $J_{min}=0.005$.  This minimal value is taken since  the smaller intensities are  hardly observable in the experiment}. The size $N_c$ of the cluster of MQ coherences does evolve, which is demonstrated in 
Fig.\ref{Fig:exp13}. This evolution is a consequence of the fact that the rate of the decay increases with the increase in the order of
MQ NMR coherences.  We see also that the rate of decrease of the { coherence} cluster size  increases with the increase in $N$. 
The described experiment may be used in order to prepare the { coherence} clusters of desirable size varying the duration of the evolution period.

\subsection{The decay of MQ NMR coherences in MQ NMR experiments with the modified preparation period}
\label{Section:second}
{ It is very important to investigate the degradation of quantum superposition states.}
 MQ NMR experiments \cite{BMGP} allow 
us to make it. To this end the modification of the preparation period of the MQ NMR experiment was suggested in 
ref.\cite{AS}. In this section we consider Eqs.(\ref{H_eff}) and (\ref{H_ev}) with  $p>0$, so that the system is governed by the general Hamiltonian $H_{eff}$ during the preparation period and by the Hamiltonian $H_{ev}=\Delta I_z$ during the evolution period. Thus the coherence decay occurs on the preparation period of  the MQ NMR experiment. 
The calculations analogous to those used for the derivation of Eq.(\ref{J_n}) yield the following expression for the intensities of MQ NMR coherences:
\begin{eqnarray}\label{J_n_p}
J_k(\tau,p)={\mbox{Tr}}\{
\tilde\rho_k(\tau,p)\rho_{-k}(\tau,p)
\},\;\;\;\tau=K\tau_c,
\end{eqnarray}
where
\begin{eqnarray}
\tilde\rho(\tau,p)=e^{-i\tau H_{eff}} I_z e^{i \tau H_{eff}}=\sum_{k} \tilde \rho_k.
\end{eqnarray}

If $p\ll 1$, then it is simple to demonstrate that the intensities vary proportionally to $p^2$. 
In fact, the Liouville equation on the preparation period   can be rewritten as follows:
\begin{eqnarray}\label{L2}
&&
i\frac{\tilde \rho( \tau)}{d\tau} =[(1-p)\bar H_{MQ} + p \bar H_{dz},\tilde \rho( \tau)].
\end{eqnarray}
Solving Eq.(\ref{L2}) by the methods of the perturbation theory \cite{G} one can obtain
\begin{eqnarray}\label{rho2}
&&
\tilde\rho(\tau)=\rho( \tau) -p \rho_1(\tau) -p^2 \rho_2(\tau),\\\nonumber
&&
\rho_1(\tau)= i\int_0^{\tau}[e^{i\bar H_{MQ}( \tau'-\tau)}(\bar H_{dz}-\\\nonumber
&&
\bar H_{MQ}) e^{-i\bar H_{MQ}( \tau'- \tau)},\rho(\tau)]d\tau',\\\nonumber
&&
\rho_2(\tau)= \int_0^{\tau}d\tau'\int_0^{\tau'}d\tau''[e^{i\bar H_{MQ}( \tau'-\tau)}(\bar H_{dz}-\\\nonumber
&&
\bar H_{MQ}) 
e^{-i\bar H_{MQ}( \tau'- \tau)},[e^{i\bar H_{MQ}( \tau''-\tau)}(\bar H_{dz}-\\\nonumber
&&
\bar H_{MQ}) 
e^{-i\bar H_{MQ}( \tau''- \tau)},\rho(\tau)]],
\end{eqnarray}
where $\rho(\tau)$ is the solution to Eq.(\ref{L}). It is evident from Eq.(\ref{rho2}) that 
\begin{eqnarray}\label{Tr2}
&&{\mbox{Tr}}\{\tilde \rho(\tau) \rho(\tau)\} =
{\mbox{Tr}}\{\rho^2( \tau)\} -p^2 A(\tau),\\\nonumber
&&
A( \tau)=\int_0^{\tau}d\tau'\int_0^{\tau'}d\tau''{\mbox{Tr}}\Big\{\\\nonumber
&&
[\rho(t),
e^{i\bar H_{MQ}( \tau'-\tau)}(\bar H_{dz}-\bar H_{MQ}) 
e^{-i\bar H_{MQ}( \tau'- \tau)}]\times
\\\nonumber
&&
[e^{i\bar H_{MQ}( \tau''-\tau)}(\bar H_{dz}-\bar H_{MQ}) 
e^{-i\bar H_{MQ}( \tau''- \tau)},\rho(\tau)]\Big\}.
\end{eqnarray} 
The behavior of the  intensities  with the increase in $p$ is defined by the sign of $A(\tau)$. 
We do not determine this sign for an arbitrary $\tau$. However,  one has for small $\tau$:
\begin{eqnarray}
A\approx -\frac{\tau^2}{2} {\mbox{Tr}}[\rho,\bar H_{dz}-\bar H_{MQ}]^2 > 0.
\end{eqnarray}
Since ${\mbox{Tr}}\{\rho^2\}$ is the sum of the intensities of MQ NMR coherences for the standard MQ NMR experiment and ${\mbox{Tr}}\{\tilde \rho(\tau) \rho(\tau)\}$ is the analogous sum, when the perturbations are taken into account one can conclude that the intensities decrease with the increase in the square of the perturbation strength at least for small $\tau$, which is confirmed below by the numerical simulations.

%%%%%%%%%%%%%%%%
\subsubsection{The numerical simulations}
\label{Section:num_second}
\begin{figure}
  \epsfig{file=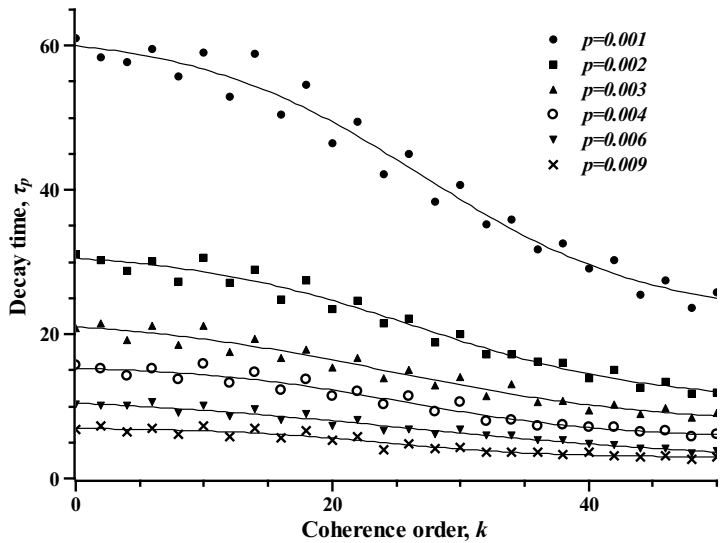
%figexp2N201
   ,scale=0.6,angle=270}
  \caption{The decay time as a function of the coherence number for spin systems with $N=201$ }
  \label{Fig:exp21}
\end{figure}

{ 
 We refer to $\bar \tau=D\tau$ as the dimensionless evolution time in this section.
Before proceed to the numerical simulations, let us underline the basic difference between the experiments considered in  Secs.\ref{Section:first} and \ref{Section:second}. The matter is  that the high frequency oscillations of MQ NMR coherences  are formed on the preparation period with the duration $\tau$, while the decay of these coherences occurs on the evolution period (with the duration $t$)   in Sec.\ref{Section:first}. For this reason we consider the intensities averaged over the parameter $\tau$ in that section. 
However, the situation is different in Sec.\ref{Section:second}, because  the decay occurs on the preparation period, so that the parameter $\tau$ is responsible for the both oscilations and decay of MQ  NMR coherences. Because of this fact, we are not able to consider the averaged intensities.  
%The numerical simulations of Eq.(\ref{J_n_p}) show that intensities are oscillating functions so that the

Instead of this, we relate the decay time $\tau_p(k)$ of the intensity $J_k(\bar\tau)$  with the decay times $\tau_p^{\pm}(k)$ of its envelopes. Here the subscript $p$ indicates that  the parameter $\tau_p$ depends on the value of $p$ in the Hamiltonian $H_{eff}$.   Parameters $\tau_p^{\pm}(k)$ may be  found simply by plotting the graphs of the envelopes $\hat J_k^{\pm}(\bar\tau)$ of the quickly oscillating intensity  $J_k(\bar\tau)$, 
$\hat J_k^{-}(\bar\tau)\le J_k(\bar\tau)\le \hat J_k^{+}(\bar\tau)$. Then 
the decay times of the envelopes $\tau_p^{\pm}(k)$ are the first zeros of $\hat J_k^{\pm}(\bar\tau)$ appeared after the amplitude of the $k$th intensity gets its maximal value:  
$\hat J_k^{\pm}(\tau^{env}_p(k))=0$.   All this suggests us to calculate the decay time $\tau_p(k)$  of the $k$th coherence  as 
follows. First, we have to find numerically 
 all  solutions $ \tau_p^{(i)}(k)$ ($i=1,2,\dots$) to the  equation $J_k(\bar\tau,p)|_{\bar\tau=\tau_p^{(i)}(k)}=0$, such that 
 $  \tau^{-}_p(k)<\tau_p^{(i)}(k)\le \tau^{+}_p(k)$. Let $N_{p}(k)$ be the number of such solutions. Then the decay time of the $k$th coherence intensity may be found as the averaged value of these solutions: $\tau_p(k)=\frac{1}{N_{p}(k)}\sum_{i=1}^{N_{p}(k)} \tau_p^{(i)}(k)$.} The dependence of $\tau_p(k)$ on the coherence number is shown in Fig.\ref{Fig:exp21} for $N=201$. 
It is found that this decay may be approximated as follows:
$\tau_p(k)\approx a_p+b_p\tanh( d_p-c_p k )$. Parameters $a_p$, $b_p$, $c_p$, $d_p$ have been found by the least square method, see Table I. 
 \begin{table}[!htb]
\begin{tabular}{|c|c|c|c|c|}
\hline
%{|p{0.5cm}|p{2.cm}|p{1.5cm}|p{1.5cm} |p{1.5cm}|}
p    &$a_p$  &$b_p$   &$c_p$&$d_p$\\\hline
0.001&42.0073&19.7734&0.0565&1.5240\\
0.002&21.1130&10.5523&0.0543&1.4369\\
0.003&14.9127&7.4843 &0.0472&1.1474\\
0.004&10.6510&5.0798 &0.0606&1.5382 \\
0.006&6.9864 &4.7489 &0.0358&0.9288\\
0.009&4.9630 &2.0483 &0.0778 &1.8647\\\hline
\end{tabular}
\label{Table:HPST1}
\caption{The parameters of the approximation $\tau_p(k)=a_p+b_p\tanh( d_p-c_p k )$ for the spin system with $N=201$ }
\end{table}
{ We see that  the  decay time of the high order coherence intensity  depends slightly on its order in accordance with the represented formula. This conclusion is  similar to that given in Sec.\ref{Section:num_first}, see eq.(\ref{coth}).}

\begin{figure}
  \epsfig{file=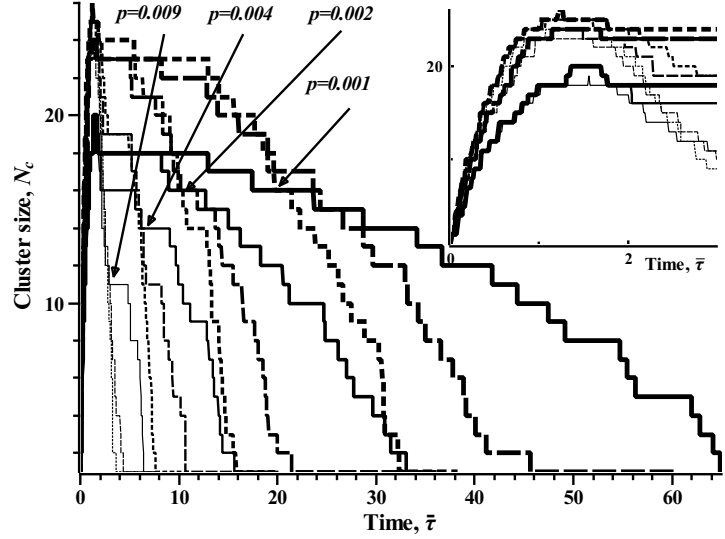
%figexp22
   ,scale=0.6,angle=270}
  \caption{The evolution of the cluster size $N_c$ for $N=201$ (solid line), $N=401$ (dashed line), $N=601$ (dotted line) and different $p$; the widths of the lines increase with the decrease of the parameter $p$. The inset shows the periods of the coherence cluster growth and decay at small times.}
  \label{Fig:exp22}
\end{figure}

Similar to Sec.\ref{Section:num_first}, we may introduce  the cluster of MQ coherences at any time moment $\bar\tau$ as
a family of such coherences that $\hat J^+_k(\bar \tau)\ge J_{min} =0.005$.
% which have not decayed by the time moment $\bar \tau$, i.e. the $k$th coherence belongs to the cluster if   %$\bar\tau<\tau_p(k) $.
 Evolution of the cluster size $N_c(\bar\tau)$ is shown in Fig.\ref{Fig:exp22}
for different $N$ and $p$. 
{ We see  that the results of our simulations agree qualitatively with the experimental ones obtained in \cite{AS}. Namely, there is  the period of the coherence cluster  growth $0\le \bar\tau \lesssim 1.5$ \cite{AS,BMGP} and the period of the cluster decay, $\bar \tau \gtrsim 1.5$. Fig.\ref{Fig:exp22} demonstrates that the cluster size gets its maximal value at  the time moment $\bar\tau\approx 1.5$, which is slightly dependent on the both parameters $N$ and $p$.  This confirms our assumptions that all spins become embedded in the cluster during the  time interval $\tau\sim 1/D$, or $\bar \tau \sim 1$. This feature of the  cluster growth in the system of equivalent spins is different from that  in solids \cite{AS}. }
Fig.\ref{Fig:exp22}  demonstrates also that the maximal  size of the cluster increases with the increase in $N$ and slightly decreases with the increase in $p$.
 The rate of the cluster decay increases with the increase in both $N$ and $p$. 
%This is the consequence of the fact that the rate of the  MQ coherence decay increases with the increase in the %coherence order,  parameter $p$ and  spin number $N$.
%, similar to the Sec.\ref{Section:num_first}.

{
Comparison of Figs.\ref{Fig:exp13} and \ref{Fig:exp22}
shows that the  case of the modified preparation period is more flexible in preparation of the coherence clusters with the desirable size because of the parameter $p$ which does not appear in Sec.\ref{Section:first}.  }

 %%%%%%%%%%%%%
\section{The conservation law in the model of the dipolar relaxation of MQ NMR coherences}
\label{Section:discussion}
{  It is worth to emphasize that the appearance of MQ NMR coherences and their relaxation are determined by the same DDI, which is valid  in the models both suggested in \cite{KS1,KS2,AS} and  considered in the previous sections. This leads to some peculiarities of the relaxation process.} We show that the sum of areas of the signals of MQ NMR coherences in the frequency domain is not changed in the relaxation process although their maximal amplitudes decrease. For the sake of simplicity, we turn to the case considered in Sec.\ref{Section:first}, where the decay occurs on the evolution period. Then the intensities of MQ NMR coherences are determined by Eq.(\ref{J_n}). Performing the Fourier transform of the intensities  $J_k(\tau,t)$ of Eq.(\ref{J_n}) over the time $t$ of the evolution period (we suppose that $J_k(\tau,t)=0$ for $t<0$ and $t>T$, where $T$ is the duration of the evolution period) 
\begin{eqnarray}\label{F1}
{\cal{J}}_k(\tau,\omega)=\frac{1}{2\pi}\int_0^T J_k(\tau,t) e^{-i\omega t} dt,
\end{eqnarray}
one can find that the area $A_k(\tau)$ under ${\cal{J}}_k(\tau,\omega)$ in the frequency domain is 
\begin{eqnarray}\label{F2}
&&
A_k(\tau)=\frac{1}{2\pi}\int_{-\infty}^\infty {\cal{J}}_k(\tau,\omega) d\omega =\\\nonumber
&&
\frac{1}{2\pi} \int_{-\infty}^\infty d\omega  \int_{0}^T J_k(\tau,t) e^{-i\omega t} d t=\\\nonumber
&&
\frac{1}{2\pi}\int_0^TJ_k(\tau,t) dt  \int_{-\infty}^\infty e^{-i\omega t} d\omega = \frac{1}{2}J_k(\tau,0). 
\end{eqnarray}
%An infinite upper limit in the integral  of Eq.(\ref{F1}) is justified by the exponential decay of many-spin correlators %at long times, which is usual in spin dynamics \cite{G}.
{ Then the sum of the areas $A_k$ for all MQ NMR coherences  can be expressed as follows:
\begin{eqnarray}\label{A}
\sum_k A_k(\tau)=\frac{1}{2}\sum_k J_k(\tau,0)
%=\frac{1}{2}
.
\end{eqnarray}
However, it is known  that $\sum_k J_k(\tau,0)=1$ \cite{LHG}. 
Thus, 
eq.(\ref{A}) means that the areas $A_k(\tau)$ are redistributing during the relaxation process so that  their sum is conserved.}

Similarly, replacing $t$ with $p$ and $T$ with $1$ in Eqs.(\ref{F1}) and (\ref{F2}) one derives the same conservation law for  MQ NMR experiment of Sec.\ref{Section:second}.
 
{  The results of this section demonstrate some peculiarities of the used relaxation model.}

%%%%%%%%%%%%
\section{Conclusions}
\label{Section:conclusion}
{ Using the numerical methods describing the spin dynamics in large systems of equivalent spins \cite{DFFZ1,DFFZ2}, we study the decay of MQ NMR coherences in such systems.} This decay is caused by the Hamiltonian $\bar H_{dz}$ appearing either on the preparation or evolution period of the MQ NMR experiment.  Numerical simulations are performed for the systems consisting of 200-600 spins. It is found that the relaxation rate of MQ NMR coherences from the highly correlated spin states increases with the  increase in both the MQ NMR order and the number of spins.  The dependence of the relaxation time of  MQ NMR coherences on the perturbation strength $p$, appearing on the preparation period, is also investigated. { We emphasize that  the used model \cite{KS1,KS2,AS}  is the first one for the experimental investigation of the relaxation of the correlated spin clusters of the large size.}

{ It is worth to note that the evolution   of the intensities of MQ NMR coherences in the system of equivalent spins  is accompanied by the reversion phenomena.   Such phenomena were studied both experimentally and numerically 
\cite{RSOPL,SLAC} and the decoherence was considered as the decay of the Loschmidt echo.  
The reversion phenomena are not considered in this paper.}

All numerical simulations have been performed using the resources of the Joint Supercomputer Center (JSCC) of the Russian Academy of Sciences. Authors thank the anonymous referee for the valuable remarks. The work was supported by the Program of the Presidium of Russian Academy of Sciences No.21 " Foundations of fundamental investigations of nanotechnologies and nanomaterials".

\end{document}